\documentclass[
 preprint,
superscriptaddress,
  aps,%linenumbers,
superscriptaddress,
 amsmath,amssymb,
 longbibliography
]{revtex4-1}

\usepackage{nicefrac, xfrac}
\usepackage{graphicx}% Include figure files
\usepackage{dcolumn}% Align table columns on decimal point
\usepackage{bm}% bold math
\usepackage{amsmath,amssymb,amstext,amsthm}
\usepackage{braket}
\usepackage{bbold}
\usepackage{bbm}
\usepackage{xcolor}
\usepackage{ gensymb }
\usepackage{soul}
\usepackage{multirow}

\begin{document}

%\preprint{APS/123-QED}
%\linenumbers
\setlength{\abovedisplayskip}{1pt}

%\title{K-space model and contrast optimization for phase-grating moir\'e interferometry }
%\title{A k-space model for phase-grating moir\'e interferometry }
\title{Cone beam neutron interferometry: from modeling to applications}
\author{D. Sarenac}
\email{dsarenac@uwaterloo.ca}
\affiliation{Institute for Quantum Computing, University of Waterloo,  Waterloo, ON, Canada, N2L3G1}
\affiliation{Department of Physics, University at Buffalo, State University of New York, Buffalo, New York 14260, USA}

\author{G. Gorbet} 
\affiliation{Department of Physics, University of Waterloo, Waterloo, ON, Canada, N2L3G1}

\author{C. Kapahi} 
\affiliation{Institute for Quantum Computing, University of Waterloo,  Waterloo, ON, Canada, N2L3G1}
\affiliation{Department of Physics, University of Waterloo, Waterloo, ON, Canada, N2L3G1}

\author{Charles W. Clark}
\affiliation{Joint Quantum Institute, National Institute of Standards and Technology and University of Maryland, College Park, Maryland 20742, USA}

\author{D. G. Cory}
\affiliation{Institute for Quantum Computing, University of Waterloo,  Waterloo, ON, Canada, N2L3G1}
\affiliation{Department of Chemistry, University of Waterloo, Waterloo, ON, Canada, N2L3G1}

\author{H. Ekinci} 
\affiliation{Institute for Quantum Computing, University of Waterloo,  Waterloo, ON, Canada, N2L3G1}

\author{S. Fangzhou} 
\affiliation{J-PARC Center, Japan Atomic Energy Agency (JAEA), 2-4 Shirakata, Tokai, Ibaraki 319-1195, Japan}

\author{M. E. Henderson} 
\affiliation{Institute for Quantum Computing, University of Waterloo,  Waterloo, ON, Canada, N2L3G1}
\affiliation{Department of Physics, University of Waterloo, Waterloo, ON, Canada, N2L3G1}

\author{M. G. Huber}
\affiliation{National Institute of Standards and Technology, Gaithersburg, Maryland 20899, USA}

\author{D. Hussey}
\affiliation{National Institute of Standards and Technology, Gaithersburg, Maryland 20899, USA}

\author{P. A. Kienzle}
\affiliation{National Institute of Standards and Technology, Gaithersburg, Maryland 20899, USA}

\author{R. Serrat} 
\affiliation{Institute for Quantum Computing, University of Waterloo,  Waterloo, ON, Canada, N2L3G1}
\affiliation{Centre de Formació Interdisciplinària Superior - Universitat Politècnica de Catalunya, 08028 Barcelona, Spain}

\author{J. D. Parker} 
\affiliation{J-PARC Center, Japan Atomic Energy Agency (JAEA), 2-4 Shirakata, Tokai, Ibaraki 319-1195, Japan}

\author{T. Shinohara} 
\affiliation{J-PARC Center, Japan Atomic Energy Agency (JAEA), 2-4 Shirakata, Tokai, Ibaraki 319-1195, Japan}

\author{D. A. Pushin}
\email{dmitry.pushin@uwaterloo.ca}
\affiliation{Institute for Quantum Computing, University of Waterloo,  Waterloo, ON, Canada, N2L3G1}
\affiliation{Department of Physics, University of Waterloo, Waterloo, ON, Canada, N2L3G1}

\date{\today}% It is always \today, today,

%\begin{abstract}
%\end{abstract}

\pacs{Valid PACS appear here}

\begin{abstract}
Phase-grating moir\'e interferometers (PGMIs) have emerged as promising candidates for the next generation of neutron interferometry, enabling the use of a polychromatic beam and manifesting interference patterns that can be directly imaged by existing neutron cameras. However, the modeling of the various PGMI configurations is limited to cumbersome numerical calculations and backward propagation models which often do not enable one to explore the setup parameters. Here we generalize the Fresnel scaling theorem to introduce a k-space model for PGMI setups illuminated by a cone beam, thus enabling an intuitive forward propagation model for a wide range of parameters. The interference manifested by a PGMI is shown to be a special case of the Talbot effect, and the optimal fringe visibility is shown to occur at the moir\'e location of the Talbot distances. We derive analytical expressions for the contrast and the propagating intensity profiles in various conditions, and analyze the behaviour of the dark-field imaging signal when considering sample characterization. The model’s predictions are compared to experimental measurements and good agreement is found between them. Lastly, we propose and experimentally verify a method to recover contrast at typically inaccessible PGMI autocorrelation lengths. The presented work provides a toolbox for analyzing and understanding existing PGMI setups and their future applications, for example extensions to two-dimensional PGMIs and characterization of samples with non-trivial structures.
\end{abstract}
\maketitle

\section{Introduction}

Outstanding issues in fundamental physics and ongoing advances in material science have created a need for the development of novel interferometry techniques and characterization tools. Neutrons are a powerful probe of nature and materials, as their nanometer-sized wavelengths and electric neutrality enable unique scattering capabilities that are complementary to x-rays and electrons~\cite{klein1983neutron,abele2008neutron,willis2017experimental}. The development and deployment of neutron interferometry setups and diffraction components remains a vibrant research field~\cite{oam,decoherence,holography,Klepp_2014}. 

Traditional Mach-Zehnder designs of neutron interferometers enjoyed a long record of success in performing fundamental tests of nature and exploring neutron interactions~\cite{ni_book2ed}. However, the stringent requirements on alignment, environmental isolation, and narrow wavelength acceptance~\cite{Arif_1994_VibrMonitCont,saggu2016decoupling,pushin2015neutron} have shifted the focus onto grating-based designs that are capable of working in the full field of the neutron beam, and require a relatively low amount of isolation~\cite{Clauser1994,Pfeiffer2006,croninnano,daviddifferential, Cronin_2009_RMP,chapman1995near,lau1948}. These emerging setups employ the ``Talbot effect'' which describes the self-imaging that occurs after a wave passes through a periodic structure~\cite{TalbotEffect}. For example, passing neutrons with wavelength $\lambda$ through a grating with period $p_G$ induces a self-image of the grating profile after the neutrons propagate a distance of $p_G^2/\lambda$. A shortcoming in these neutron setups is the fact that they require the placement of an absorption-grating at the imaging plane in order to observe the self-image. The self-image from micron-sized gratings is not detectable via typical neutron cameras with resolution $\approx 10~\mu$m. 

The newest form of neutron interferometry setups are the phase-grating moir\'e interferometers (PGMIs)  whose optimal fringe visibility occurs with periods on the millimeter scale, thus enabling direct detection with a neutron camera without the need for an absorption grating~\cite{twogratings,hussey2016demonstration,sarenac2018three,brooks2018neutron,brooks2017neutron}. The PGMIs are composed of exclusively phase-gratings and work in the full field of a cone beam~\cite{miao2016universal}. However, the rather large span in the length scales makes the simulation of relevant coherent phenomena with cone beam illumination particularly difficult: the typical values include wavelength distributions that peak in the angstrom range, diffraction features on the micrometer scales, and propagation distances that can span several meters. The work presented here addresses this problem by introducing a forward propagation model that can be used to explore complicated geometries and various setup parameters. The model is derived by extending the Fresnel scaling theorem~\cite{paganin2006coherent} to the illumination of multiple objects, and is similar to the k-space formalism in magnetic resonance~\cite{sodickson1998generalized,cory1989chemical,ramsey1950molecular}.

The manuscript is structured as follows: Section II provides a brief introduction to self-imaging and the Talbot effect. Section III provides the derivation of the k-space model for simulating the cone beam illumination of multiple objects. Analytical equations are provided allowing calculation of the moir\'e interference pattern and interferometer contrast for different geometries. In Section IV we show good agreement with previously published PGMI data, for both two-PGMI and three-PGMI setups. In the following two subsections we analyze the use of source gratings and the dark field imaging (DFI) technique to characterize microstructures. Lastly, we predict and experimentally verify that it is possible to recover fringe visibility at typically inaccessible interferometer autocorrelation lengths by varying the location of the phase-gratings in the setup.

\section{Phase-Gratings and the Talbot Effect} 

After a neutron passes though a material of thickness $h$ it accumulates a phase shift of:

\begin{align}
	\alpha = \frac{2\pi (1-n) h}{\lambda}=Nb_c\lambda h,
	\label{eqn:phase}
\end{align}

\noindent where $n$ is the material's index of refraction and $Nb_c$ is the scattering length density. It follows that a binary phase-grating with a 50~\% duty-cycle, an offset of $x_1$, period $p_G$, and height $h$, induces a periodic phase shift over the wavefront: 

\begin{align}
	f(x)=\frac{\alpha}{2} \text{sgn}\left(\text{sin}\left[\frac{2\pi }{p_G}(x-x_1)\right]\right).
	\label{eqn:gratingProfile}
\end{align}

We first take the case of a collimated beam that is well-approximated by $k_x=0$, $k_z=|k_r|=k_0=2\pi/\lambda$. Upon propagating through the phase-grating the transverse  momentum spectra is given by $P(k_x)=|\Psi(k_x)|^2$ where the transverse momentum wavefunction is given by:

\begin{align}
	\Psi(k_x) =\Psi_j(k_x)*\mathcal{F}\{e^{-i f(x)}\},
	\label{eqn:momentumPsi}
\end{align}

\noindent where 

\begin{align}
	\mathcal{F}\{e^{-if(x)}\} = \delta(k_x)\cos\left(\frac{\alpha}{2}\right) 
	+  \sin\left(\frac{\alpha}{2}\right) \sum_{m} \frac{2e^{-ik_Gmx_1}}{\pi m}\delta\left(k_x-mk_G\right),
	\label{eqn:gratingTerms}
\end{align}

\noindent $k_G=2\pi/p_G$ is the grating wave vector, $m=...-3,-1,1,3...$ are the non-zero diffraction orders from a binary phase-grating with a 50~\% duty-cycle, $\Psi_j(k_x)$ is the incoming momentum wavefunction, $\delta()$ is the delta function, $\mathcal{F}\{\}$ is the Fourier transform, and $*$ is the convolution operator.  The first term in Eq.~\ref{eqn:gratingTerms} is the zeroth diffraction order and the second term is the sum over the higher odd orders. A typical diffraction order spectra obtained from Eq.~\ref{eqn:gratingTerms} is depicted on Fig.~\ref{fig:diffractionSpectra}a, and can be extended to multiple gratings as shown in Fig.~\ref{fig:diffractionSpectra}b \& c. The amplitude terms dictate that for a phase-grating with $\alpha=\pi$ the zeroth order is suppressed, while $\alpha=\pi/2$ prepares a state which contains an equal amount of the zeroth order and all of the higher orders combined. 

To determine the transverse intensity profile as the beam propagates after the phase-grating we must take into account the coupling between the momentum in the transverse and propagation directions. From conservation of momentum we can infer that the phase-grating effectively does a rotation in k-space for each diffraction order and that Eq.~\ref{eqn:gratingTerms} represents the projection of the wave vector $(k_r)$ onto the transverse coordinate. Therefore, the $m^{\text{th}}$ diffraction order has a transverse wave vector of $mk_G$ and a longitudinal wave vector of:

\begin{figure*}
    \centering\includegraphics[width=\linewidth]{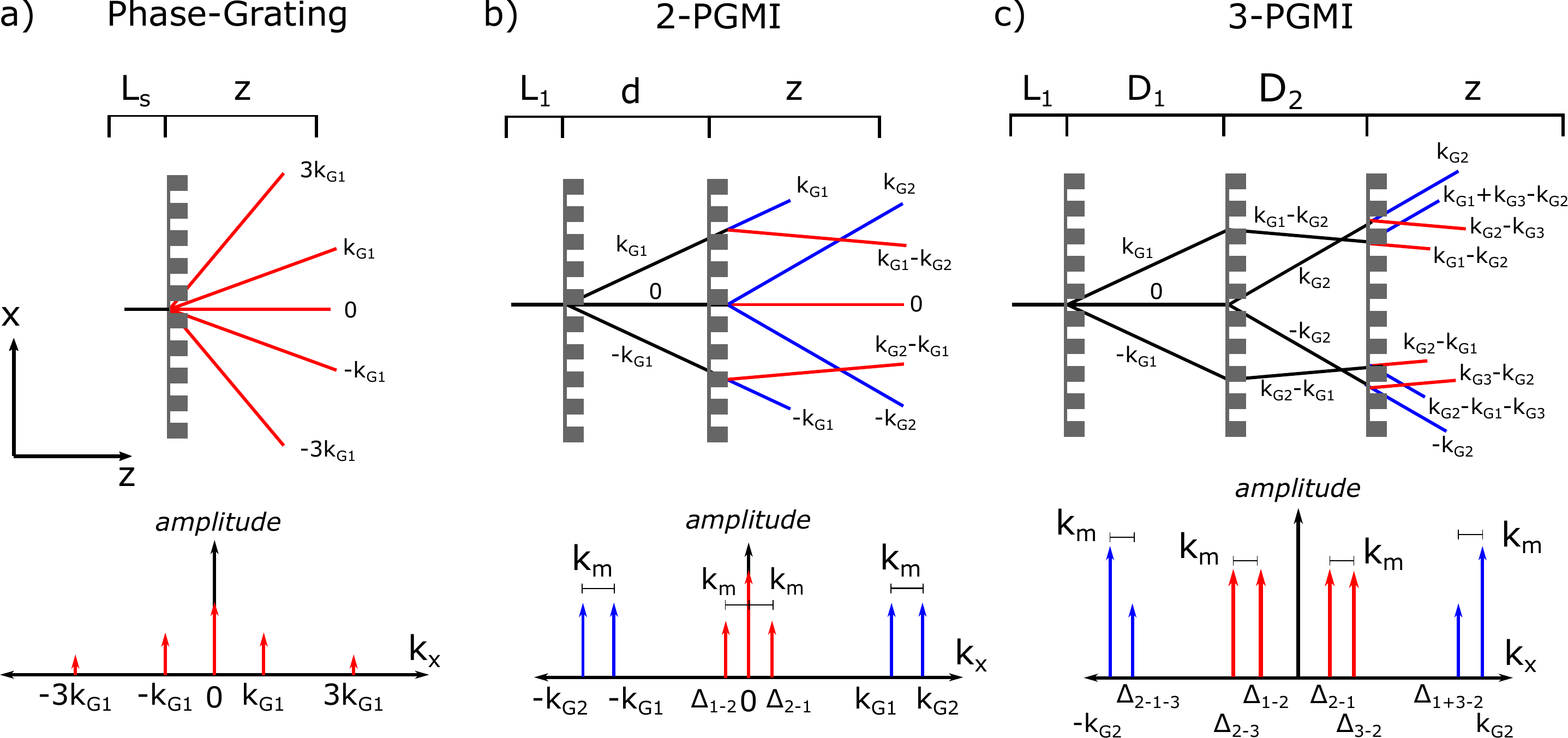}
    \caption{Central section of the diffraction spectra in phase-grating interferometers for a) a single binary $\pi/2$ phase-grating with 50~\% duty cycle b) two phase-grating moir\'e interferometer c) three phase-grating moir\'e interferometer. The amplitude of the diffraction orders is dependent on the induced phase shift, in these examples $\alpha=\pi/2$, except for the middle grating of the three PGMI where $\alpha=\pi$. The wave vector groups that provide the main contribution to a moir\'e pattern in two-PGMI and three-PGMI are highlighted in red and blue.  
    }
    \label{fig:diffractionSpectra}
\end{figure*}

\begin{align}
	\beta_m=\sqrt{k^2_0-(mk_G)^2}\approx k_0\left(1-\frac{m^2k_G^2}{2k^2_0}\right)=\frac{2\pi}{\lambda}\left(1-\frac{m^2\lambda^2}{2p^2_G}\right)
	\label{eqn:betam}
\end{align}

\noindent where the right-hand side is obtained via the paraxial approximation. The total momentum wavefunction then becomes:

\begin{align}
	\Psi(k_x,k_z) =\delta(k_x,k_z-k_0)\cos\left(\frac{\alpha}{2}\right) 
	+  \sin\left(\frac{\alpha}{2}\right) \sum_{m} \frac{2e^{-ik_Gmx_1}}{\pi m}\delta\left(k_x-mk_G,k_z-\beta_m\right)
	\label{eqn:momentumXZPsi}
\end{align}

\noindent and the intensity for any point after the phase-grating can be computed from: 

\begin{align}
I(x,z)=\left|\mathcal{F}^{-1}\{\Psi(k_x,k_z)\}\right|^2.
\label{eqn:Intensity}
\end{align}

Using the equations above, we can determine the intensity profile after the phase-grating:

\begin{widetext}
\begin{align}
	I(x,z)=\left|\cos\left(\frac{\alpha}{2}\right) 
	+ \sin\left(\frac{\alpha}{2}\right) \sum_m \frac{2}{\pi m} e^{ -i\left(\frac{m^2k_G^2}{2k_0}\right)z}e^{i mk_G (x-x_1)}\right|^2
	\label{eqn:TalbotIntensity}
\end{align}
\end{widetext}

Here we can note that the phase shifts proportional to $m^2$ ($m$) manifest themselves in the intensity profile as a translation along the propagation (transverse) direction. The latter can be obtained by translating the grating along the transverse direction and thus varying $x_1$. This is commonly referred to as ``phase-stepping''.

The Talbot effect, or self-imaging, is a near-field phenomenon that is observed after periodic structures are illuminated by waves~\cite{TalbotEffect}. It arises from the interference of the diffraction orders before they propagate into the far-field and become spatially separated. For special values of $z$ we can retrieve the original wavefunction, and thereby achieve ``self-imaging''. Substituting $z=4\pi k_0/k^2_G=2p_G^2/\lambda$ yields $e^{i2\pi m^2}=1$ for all integer values of m, and hence the result is equivalent to setting $z=0$. While setting $z=\pi k_0/k^2_G=p_G^2/\lambda$ yields $e^{i\pi m^2}=-1$ for all integer values of m, and the only difference is that there is a minus sign in front of the diffraction order sum. This represents the equation of a phase-grating at $z=0$ that has been translated by half a period, as $e^{-imk_Gx_1}=-1$ for all $m$ when $x_1=p_G/2$. The first and second distances are known as the Talbot $(z_T=2p_G^2/\lambda)$ and half Talbot distance $(\nicefrac{z_T}{2}=p_G^2/\lambda)$, and they are displayed in Fig.~\ref{fig:TalbotandMoireIntensity}a. 

\section{k-space Model for Cone Beams}

The Fresnel scaling theorem states that the intensity obtained with cone-beam illumination of an object can be modeled as a scaled parallel-beam illumination~\cite{paganin2006coherent}. The theorem dictates a geometrical magnification to the intensity profile as follows: to compute the intensity profile $\text{I}(x,z)$ after a propagation of distance $z$ following a previous propagation of distance $z_0$ we can scale the transverse and longitudinal coordinates as $(x\rightarrow x/M,z\rightarrow z/M)$ where:

\begin{align}
\text{M}=(1 + z/z_0).
\label{eqn:Magnification}
\end{align}

Here we introduce a k-space model for PGMI setups by extending the Fresnel scaling theorem to the illumination of multiple objects that are spatially separated. In order to do that, we have to consider the geometrical expansion of the wavefunction $\Psi(k_x,k_z)$ as the cone beam propagates. The model is based on the following set of postulates:

\begin{enumerate}
    \item The geometrical scaling of the spatial coordinates in the Fresnel scaling theorem can equivalently be interpreted as an input wave whose wave vector is being scaled with propagation, $k_0\rightarrow k_0/M$. That is, $e^{ik_0(r/M)}=e^{i(k_0/M)r}$. It follows that the transverse and longitudinal wave vectors are being scaled as well: $k_x\rightarrow k_x/M$ and $k_z\rightarrow k_z/M$; while the angle of propagation remains constant with propagation: $\theta=\sin^{-1}(k_x/k_0)=\sin^{-1}(Mk_x/Mk_0)$. 
    \item Each new object with a spatial phase profile of $f(x)$, introduces a convolution between $\mathcal{F}\{e^{if(x)}\}$ and the incoming momentum-space wavefunction. That is, each diffraction order of $e^{if(x)}$ rotates each incoming wave vector to point along a new direction. The previous section outlines this for the case of a single phase-grating. We can observe that as a consequence of postulate 1, in a system consisting of two identical phase gratings that are spatially separated, the second phase-grating will produce a different amount of rotation compared to the first. Note that in the case of attenuation objects the convolution kernel becomes $\mathcal{F}\{f(x)\}$.
    \item A propagation of distance $z$ induces a phase shift of $e^{ik_z z/M}$ onto each magnified diffraction order. 
    \item A new $z$-coordinate ($z'$) is defined after each new object, where $z'=z/M$ is the scaled old coordinate ($z$) and $z'=0$ is set at the location of the new object. This extra scaling factor is necessary because the induced phase due to propagation (postulate 3) is not linear with $z$  because the magnification is coupled to the $z$ coordinate. Take for instance the propagation of a distance $L_1$ followed by propagation of a distance $L_2$, resulting in a phase of $e^{ik_z L_1/(1+L_1/L_0)}e^{ik_z L_2/[(1+L_1/L_0)(1+L_2/(L_0+L_1)]}$. This is not equivalent to $e^{ik_z (L_1+L_2)/(1+(L_1+L_2)/L_0)}$ which would  result from considering a direct propagation distance of $L_1+L_2$. This behaviour is unlike the case of simple plane waves where $e^{ik_z L_1}e^{ik_z L_2}=e^{ik_z (L_1+ L_2)}$. $z'$ rectifies this whereby $e^{ik_z L_1/(1+L_1/L_0)}e^{ik_zL_2/([1+L_2/(L_0+L_1)][1+L_1/L_0]^2)}=e^{ik_z (L_1+L_2)/(1+(L_1+L_2)/L_0)}$. %and thus allows for the use of the identity operator $\delta(k_x)$.
    \item The intensity profile after the final object is computed according to Eq.~\ref{eqn:Intensity}: I$(x,z)=\left|\mathcal{F}^{-1}\{\Psi(k_x,k_z)\}\right|^2.$
    \item The ``contrast'' or ``visibility'' of a particular wave vector $(k_M)$ is computed by: $V=2|H(k_M)/H(0)|$ where $H(k_x)$ is the autocorrelation function of $\Psi(k_x)$.
    
\end{enumerate}

The above postulates allow us to compactly implement a model for a general setup consisting of $q$ objects with periodic structures, where the first object is a distance of $L_1$ from the slit, the second object is a distance of $L_2$ from the first object, and so on. For an implementation example with the two-PGMI and the corresponding pseudo code, see Appendix A. The intensity profile after the last object is given by:

\begin{widetext}
\begin{align}
	I(x,z)=\bigg|&\sum_{m_1,m_2...m_q} (a_{m_1}a_{m_2}...a_{m_q}) \\\nonumber &e^{i\frac{k_0 L_2}{M_1}  \cos\theta_{m_1}}e^{i\frac{k_0 L_3}{M_1^2M_2}  \cos\theta_{m_1,m_2}}...e^{i\frac{k_0 L_{q}}{M_1^2M_2^2...M_{q-1}}  \cos\theta_{m_1,m_2...m_{q-1}}}\\\nonumber &e^{i\frac{1}{M_1 M_2...M_q} \frac{1}{1+z/(L_1+L_2+L_3+...+L_q)} k_0(\frac{z}{M_1 M_2...M_{q-1}} \cos\theta_{m_1,m_2...m_q}+x\sin\theta_{m_1,m_2...m_q})}\bigg|^2
\end{align}
\end{widetext}

\noindent where $z=0$ is the location of the last object, $\sum_{m_j} a_{m_j}$ are the Fourier coefficients of the $j^{th}$ object, $M_j=1+L_{j+1}/(L_1+L_2+...+L_j)$ is the scaling factor between object $j$ and $j+1$, and $\theta_{m_1,m_2...,m_j}=\sin^{-1}[(M_1M_2...M_{j-1}k_{G1}+M_1M_2...M_{j-2}k_{G2}+k_{Gi})/(M_1M_2...M_{j-1} k_0)]$ are the diffraction angles after the object $j$  with wave vector $k_{Gi}$.  

After a single object located at $L_1$ from the slit, the intensity at any point (x,z) after the object is given by: 

\begin{widetext}
\begin{align}
	I(x,z)=\left|\sum_m a_m e^{i\frac{1}{1+z/L_1} k_0(z \cos\theta_m+x\sin\theta_m)}\right|^2
	\label{eqn:TalbotIntensity3}
\end{align}
\end{widetext}

\noindent which in the small angle approximation for a 50~\% duty-cycle phase-grating reduces to

\begin{figure*}
    \centering\includegraphics[width=\linewidth]{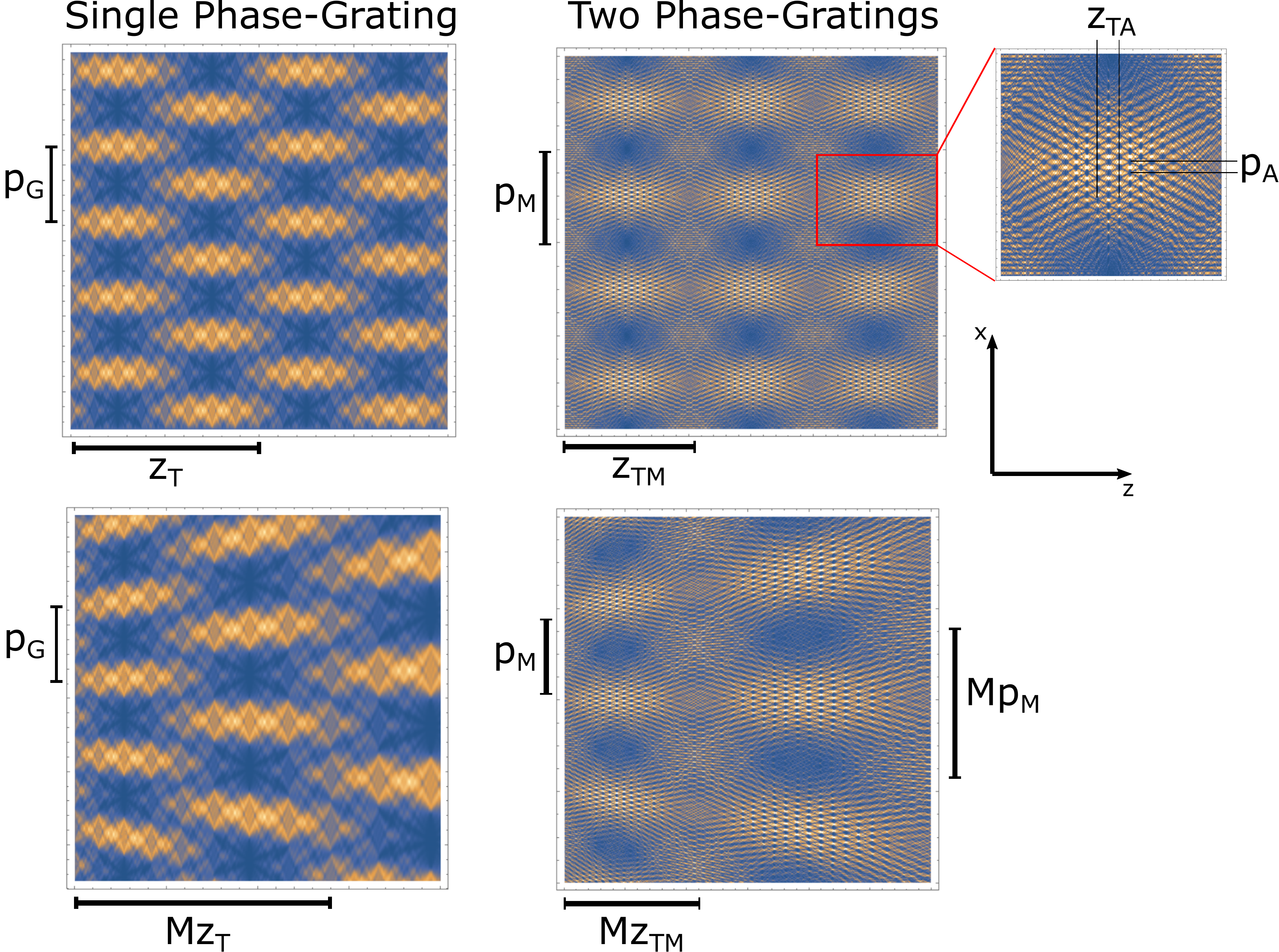}
    \caption{ The top row considers the propagating intensity profile in a parallel beam geometry after a single $\pi/2$ phase-grating with period $p_G$ and after two $\pi/2$ phase-gratings with slightly different periods manifesting a moir\'e period of $p_M$. The single phase-grating manifests the well-known Talbot effect whereby a self-image is obtain after a propagation of $z=2p_G^2/\lambda\equiv z_T$. The interference between the diffraction orders of two phase-gratings with slightly different periods gives rise to the moir\'e period $(p_M)$, Talbot-moir\'e distance $(z_{TM})$, average period $(p_A)$, and average Talbot distance $(z_{TA})$. The bottom row considers a cone beam geometry whereby the moir\'e arises even when the two phase-gratings have the same period. $p_M$ and $z_{TM}$ are typically observed in a neutron PGMI setup, whereas the higher frequency interference effects (shown in the inset) are not resolved. Lastly, we can note that in the case of the cone beam illumination the period of the interference patterns as well as the self-imaging distances are continuously magnified by a factor $M=(1+z/z_0)$, where $z_0$ is the distance from the source to the last grating.
    }
    \label{fig:TalbotandMoireIntensity}
\end{figure*}

\begin{widetext}
\begin{align}
	I(x,z)=\left|\cos\left(\frac{\alpha}{2}\right) 
	+ \sin\left(\frac{\alpha}{2}\right) \sum_m \frac{2}{\pi m} e^{-i mk_G x_1}e^{ -i\left(\frac{m^2k_G^2}{2k_0(1+z/L_1)}\right)z}e^{i \frac{mk_G}{1+z/L_1}x }\right|^2
	\label{eqn:TalbotIntensity4}
\end{align}
\end{widetext}

\noindent where $\alpha$ is given by Eq.~\ref{eqn:phase} and the $a_m$ terms are given in Eq.~\ref{eqn:gratingTerms}. Note that Eq.~\ref{eqn:TalbotIntensity4} is identical to Eq.~\ref{eqn:TalbotIntensity} with the added feature that the wave vectors are being scaled with propagation $k_x\rightarrow k_x/(1+z/L_1)$ and $k_z\rightarrow k_z/(1+z/L_1)$. It follows that the size of the ``self-image'' as well as the Talbot distance increase with propagation. This is shown in the bottom row of Fig.~\ref{fig:TalbotandMoireIntensity}a.

Considering the addition of a second object at a distance of $L_2$ from the first object, the intensity at any location (x,z) after the second object is: 

\begin{widetext}
\begin{align}
	I(x,z)=\left|\sum_{m,n} a_m a_n e^{i\frac{k_0}{M_1}L_2 \cos\theta_m} e^{i\frac{1}{M_1}\frac{1}{1+z/(L_1+L_2)} k_0(\frac{z}{M_1} \cos\theta_{m,n}+x\sin\theta_{m,n})}\right|^2
	\label{eqn:TalbotIntensity5}
\end{align}
\end{widetext}

\noindent where $M_1=1+L_2/L_1$ and $M_2=1+z/(L_1+L_2)$. In the notation of a two-PGMI let us consider a cone beam that travels through the first phase-grating located a distance of $L_1$ away from the slit, followed by free-space propagation of distance $d$, followed by the second phase-grating. In the small angle approximation we can obtain the intensity profile as the beam propagates further:

\begin{widetext}
\begin{align}
	\nonumber I(x,z)=&\bigg|\mathcal{C}_1\mathcal{C}_2
	+\mathcal{C}_1\mathcal{S}_2 \sum_{n} \frac{2}{\pi n} e^{i n k_{G} (\frac{x}{M_2}-x_2)}e^{-i \frac{n^2 k^2_G }{2k_0M_2}z}
	\\\nonumber +&\mathcal{S}_1\mathcal{C}_2  \sum_{m} \frac{2}{\pi m} e^{i \frac{m^2k^2_G d}{2k_0M_1}}e^{im k_{G}\frac{x}{M_1M_2}}e^{-i \frac{m^2k^2_G}{2k_0M^2_1M_2}z}\\ +& \mathcal{S}_1\mathcal{S}_2 \sum_{n,m} \frac{4}{\pi^2 m n} e^{i \frac{m^2k^2_G d}{2k_0M_1}}e^{-i nk_{G}x_2}e^{i (\frac{mk_{G}}{M_1}+nk_{G})\frac{x}{M_2}}e^{-i \frac{(m/M_1+n)^2 k^2_G }{2M_2k_0}z}\bigg|^2
	\label{eqn:2PGMIconeIntensity}
\end{align}
\end{widetext}

\noindent where $\mathcal{S}_j=\sin(\alpha_j/2)$,  $\mathcal{C}_j=\cos(\alpha_j/2)$. Eq.~\ref{eqn:2PGMIconeIntensity} gives an analytical expression for the spatial intensity profile after the second grating in a two-PGMI, and it is depicted in the bottom row of Fig.~\ref{fig:TalbotandMoireIntensity}b. Although one has the freedom to include an arbitrary number of diffraction orders, the seventh orders and above have negligible influence given that the amplitude of the diffraction orders scale with $1/m$. 

After three objects the intensity at any location (x,z) after the third object is 

\begin{widetext}
\begin{align}\nonumber
	I(x,z)=\bigg|\sum_{m,n,\ell} a_m a_n a_\ell &e^{i\frac{k_0}{M_1} L_2 \cos\theta_m}e^{i\frac{k_0 L_3}{M_1^2 M_2} \cos\theta_{m,n}}\\ &e^{i\frac{1}{M_1 M_2} \frac{1}{1+z/(L_1+L_2+L_3)} k_0(\frac{z}{M_1 M_2} \cos\theta_{m,n,\ell}+x\sin\theta_{m,n,\ell})}\bigg|^2
	\label{eqn:3objects}
\end{align}
\end{widetext}

Let us consider a typical three-PGMI setup with slit to first grating to second grating to third grating distances being $L_1,L_2,L_3$, respectively. In the small angle approximation for binary phase-gratings the intensity after the third grating is given by:

\begin{widetext}
\begin{align}
	\nonumber I(x&,z)=\bigg|\mathcal{C}_1\mathcal{C}_2\mathcal{C}_3+\mathcal{C}_1\mathcal{C}_2\mathcal{S}_3 \sum_{\ell} \frac{2}{\pi \ell} e^{i \ell k_{G} (\frac{x}{M_3}-x_3)}e^{i \frac{\ell^2 k^2_G }{2k_0M_3}z}
    \\\nonumber+&\mathcal{C}_1\mathcal{S}_2\mathcal{C}_3 \sum_{n} \frac{2}{\pi n} e^{i \frac{n^2k^2_G L_3}{2k_0M_2}}e^{i n k_{G} \frac{x}{M_2M_3}}e^{-i \frac{n^2 k^2_G }{2k_0M^2_2M_3}z}
	\\\nonumber +&\mathcal{S}_1\mathcal{C}_2\mathcal{C}_3  \sum_{m} \frac{2}{\pi m} e^{-i \frac{m^2k^2_G L_2}{2k_0M_1}}e^{i \frac{m^2k^2_G L_3}{2k_0M^2_1M_2}}e^{im k_{G}\frac{x}{M_1M_2M_3}}e^{-i \frac{m^2k^2_G}{2k_0M^2_1M^2_2M_3}z}
    \\\nonumber 
    +& \mathcal{S}_1\mathcal{S}_2\mathcal{C}_3 \sum_{n,m} \frac{4}{\pi^2 m n} e^{-i \frac{m^2k^2_G L_2}{2k_0M_1}}e^{-i \frac{(m/M_1+n)^2k^2_G L_3}{2k_0M_2}}e^{i (\frac{mk_{G}}{M_1}+nk_{G})\frac{x}{M_2M_3}}e^{-i \frac{(m/M_1+n)^2 k^2_G }{2k_0M^2_2M_3}z}
    \\\nonumber 
    +& \mathcal{S}_1\mathcal{C}_2\mathcal{S}_3 \sum_{m,\ell} \frac{4}{\pi^2 m \ell} e^{-i \frac{m^2k^2_G L_2}{2k_0M_1}}e^{-i \frac{m^2k^2_G L_3}{2k_0M^2_1M_2}}e^{-i \ell k_{G} x_3}e^{i (\frac{mk_{G}}{M_1M_2}+\ell k_{G})\frac{x}{M_3}}e^{-i \frac{(m/M_1+n)^2 k^2_G }{2k_0M^2_2M_3}z}
    \\\nonumber 
    +& \mathcal{C}_1\mathcal{S}_2\mathcal{S}_3 \sum_{n,\ell} \frac{4}{\pi^2  n \ell} e^{-i \frac{n^2k^2_G L_3}{2k_0M_2}}e^{-i \ell k_{G} x_3}e^{i (\frac{nk_{G}}{M_2}+\ell k_{G})\frac{x}{M_3}}e^{-i \frac{(n/M_2+\ell)^2 k^2_G }{2k_0M_3}z}
    \\\nonumber 
    +& \mathcal{S}_1\mathcal{S}_2\mathcal{S}_3 \sum_{m,n,\ell} \frac{8}{\pi^3 m  n \ell} e^{-i \frac{m^2k^2_G L_2}{2k_0M_1}}e^{i \frac{m^2k^2_G L_3}{2k_0M^2_1M_2}}e^{i \frac{n^2k^2_G L_3}{2k_0M_2}}e^{-i \frac{(m/M_1+n)^2k^2_G L_3}{2k_0M_2}}e^{-i \ell k_{G} x_3}
    \\
    &\quad\quad\quad e^{i (\frac{mk_{G}}{M_1M_2}+\frac{nk_{G}}{M_2}+\ell k_{G})\frac{x}{M_3}}e^{-i \frac{(\frac{mk_{G}}{M_1M_2}+\frac{nk_{G}}{M_2}+\ell k_{G})^2 k^2_G }{2k_0M_3}z}\bigg|^2
	\label{eqn:3PGMIconeIntensityGratings}
\end{align}
\end{widetext}

This equation is rather general, accommodating arbitrary parameters. Considering a standard three-PGMI setup consisting of $\pi/2,\pi,\pi/2$ phase gratings and a monochromatic input, half of the terms are removed as $\mathcal{C}_2=0$. 

\subsection{Effects of experiment parameters: wavelength distribution, slit width, and camera pixel size}

The observed interference pattern at the camera is built up by the incoherent sum of each wavelength contribution. Therefore, to account for wavelength spread we can simply integrate over the particular wavelength distribution. Wavelength dependence shows up in the $k_0$ term as well as the diffraction amplitude terms: $\cos(Nb_c\lambda h/2)$ and $\sin(Nb_c\lambda h/2)$.

The analysis so far has considered a cone beam from a point source. However, a typical neutron PGMI setup incorporates the use of a slit. Note that the displacement of the point source directly results in a displacement of the interference pattern at the camera. Therefore, the slit manifests an incoherent sum of displaced interference patterns. This can either be accounted for with a straightforward convolution of output intensity with a step function of slit sized width, or integrating the point source location over the slit size. 

\iffalse
For example, in the two-PGMI:

\begin{widetext}
\begin{align}
	\int_{-\infty}^\infty d\lambda \int_0^s d x_0 [p(\lambda) I_{2PGMI}(&x-x_0, z=L_2)]
	\label{eqn:2PGMIconeIntensityIntegral}
\end{align}
\end{widetext}

\noindent where $p(\lambda)$ is the wavelength distribution, and $s$ is the slit size.
\fi
The last consideration is the camera resolution. Although the typical values of the effective camera pixel size are relatively large, they are still smaller than the slit size which creates a similar averaging effect. Pixel size can be incorporated through a convolution with a rectangular function or binning the final intensity into pixel-sized intervals.

\subsection{Fringe visibility/contrast}

The ``fringe visibility'' or ``contrast'' $(V)$ of the observed intensity profile $(I)$ is a measure of interference, and it is typically defined as:

\begin{align}
	V=\frac{I_{max}-I_{min}}{I_{max}+I_{min}},
	\label{eqn:fringeVisibility}
\end{align}

The presented k-space model allows us to determine the intensity profiles at the camera. Fitting the observed interference pattern to a sinusoid with a specific spatial frequency:

\begin{align}
	I=A+B\cos(2\pi x/p_M+\phi_0),
	\label{eqn:contrastfit}
\end{align}

\noindent where $p_M$ is the period of the oscillation, and $\phi_0$ is the phase shift, the fringe visibility of the particular frequency $2\pi/p_M$ is given by $V=B/A$. In addition to computing the contrast from the fit, the contrast value could also be computed from the Fourier Transform of the intensity profile $H(k_x)$ where $V=2|H(k_M)/H(0)|$. Determining the visibility at each pixel using a combination of phase stepping (see section II) and the Fourier transform is a typical procedure.

As per the Wiener-Khinchin Theorem, the presented k-space model also allows us to calculate the contrast from the autocorrelation of $\Psi(k_x,k_z)$: 

\begin{align}
	\mathcal{F}\{I(x,z)*f_{\text{slit}}(x)*f_{\text{pixel}}(x)\}=\left[\Psi(k_x,k_z)*\Psi^*(k_x,k_z)\right]\times F_{\text{slit}}(k_x)\times F_{\text{pixel}}(k_x).
	\label{eqn:contrastfit}
\end{align}

This result is extremely powerful and makes the determination of contrast computationally straight forward for arbitrary configurations.

\section{Observations and Model applications}

\subsection{Moir\'e of the Talbot effects}

The first observation we can make is that the observed interference in a PGMI setup is the moir\'e of the Talbot effects. Consider the case of a collimated beam illuminating a system of two phase gratings $f_1(x)$ and $f_2(x)$ with different periods $p_1$ and $p_2$. It is important to note that the neutron two-PGMI is concerned with phase profiles given by $f_1(x)+f_2(x)$ that results from the sum of two independent binary phase-gratings. This is different from the conventional concept of a beat grating with a profile in the form of $\frac{\alpha}{2} \text{sgn}\left(\text{cos}\left[\frac{2\pi }{p_1}(x-x_1)\right]+\text{cos}\left[\frac{2\pi }{p_2}(x-x_1)\right]\right)$. The profile of the conventional beat gratings possesses two amplitude values while the profile of $f_1(x)+f_2(x)$ possesses three. With a collimated beam input, the moir\'e period $(p_M)$, average period $(p_A)$, Talbot-moir\'e distance $(z_{TM})$, and the average Talbot distance $(z_{TA})$ are given by:

\begin{align}
    p_{M}=\frac{p_1p_2}{p_1-p_2} \qquad \qquad  z_{TM}=\frac{z_{T1}z_{T2}}{z_{T1}-z_{T2}}=\frac{2p_1^2p_2^2}{\lambda(p_2^2-p_1^2)}
\end{align}

\begin{align}
    p_{A}=\frac{2p_1p_2}{p_1+p_2}  \qquad \qquad 
    z_{TA}=\frac{2z_{T1}z_{T2}}{z_{T1}+z_{T2}}=\frac{4p_1^2p_2^2}{\lambda(p_1^2+p_2^2)}
\end{align}

%\begin{align}
%    \frac{p_{M}}{p_A}=\frac{p_1+p_2}{2(p_2-p_1)}   \qquad \qquad
%    \frac{z_{TM}}{z_{TA}}=\frac{z_{T1}+z_{T2}}{2(z_{T2}-z_{T1})}=\frac{p_1^2+p_2^2}{2(p^2_2-p_1^2)}
%\end{align}

\noindent These parameters are depicted on  Fig~\ref{fig:TalbotandMoireIntensity}. 
\iffalse
The moir\'e period and the Talbot-moir\'e distance at the location of the second grating are given by:

\begin{align}
    p_{M}=p\frac{(d+L_1)}{d} \qquad \qquad  z_{TM}=\frac{2p^2(d+L_1)^2}{\lambda d(d+2L_1)\lambda}
    \label{eqn:zTM}
\end{align}
\fi

With cone beam illumination, the parameters get magnified as the cone beam propagates. A typical two-PGMI setup is composed of two gratings with equal periods ($p_1=p_2=p_G$) and the distance ($L$) between the slit and the camera is set. We can rewrite the relevant variables at the location of the camera as:

\begin{align}
    p_{M}=\frac{Lp_G}{d} \qquad \qquad  z_{TM}=\frac{2p_G^2L(d+L_1)}{\lambda d(d+2L_1)}
    \label{eqn:zTM}
\end{align}

\begin{align}
    p_{A}=\frac{2Lp_G}{d+2L_1}  \qquad \qquad 
    z_{TA}=\frac{4p_G^2L(d+L_1)}{\lambda (d^2+2dL_1+2L_1^2)}
\end{align}

These parameters and their magnification with propagation is depicted on Fig~\ref{fig:TalbotandMoireIntensity}. $p_M$ and $z_{TM}$ are typically observed in a neutron PGMI setup, whereas the higher frequency interference effects (shown in the inset) are not resolved. It is expected that the development of high resolution neutron cameras~\cite{hussey2017neutron} should enable the direct observation of these finer interference effects.

\subsection{Two-PGMI analysis}

Let us now consider applying the described k-space model to the previously published works on the two-PGMI. The computational simplicity of the provided equations allows us to compute the contrast of a setup without the need for typical approximations, for example neglecting the $|m|>1$ terms. We can directly determine the contrast of a two-PGMI setup while taking into account the slit size, wavelength distribution, and camera resolution.

There are three independent two-PGMI setups that were characterized at the National Institute for Standards and Technology Center for Neutron Research (NCNR)~\cite{twogratings}. The first setup labelled ``Monochromatic'' had the following parameters: grating period $p_G=2.4~\mu$m, grating height $h=9.3~\mu$m, $L_1=1.2~$m, $L=3.04~$m, $\lambda=4.4$~\AA, slit size $200~\mu$m, and camera resolution of $100~\mu$m; the second setup labelled ``Bichromatic'' had the following parameters: grating period $p_G=2.4~\mu$m, grating height $h=9.3~\mu$m, $L_1=1.73~$m, $L=3.51~$m, $1/4.2$ intensity $\lambda=2.2$~\AA~and $3.2/4.2$ intensity $\lambda=4.4$~\AA, slit size $200~\mu$m, and camera resolution of $100~\mu $m; the third setup labelled ``Polychromatic'' had the following parameters: grating period $p_G=2.4~\mu$m, grating height $h=6.1~\mu$m, $L_1=4.65~$m, $L=8.36~$m, a wavelength distribution somewhat resembling a Maxwell-Boltzmann distribution with $\lambda_c=5$~\AA~(see Fig.~\ref{fig:wavelengthdist}a), slit size $500~\mu$m, and a camera resolution of $150~\mu$m. All of the parameters are listed in Table.~\ref{table:1}.

\begin{figure*}
    \centering\includegraphics[width=.75\linewidth]{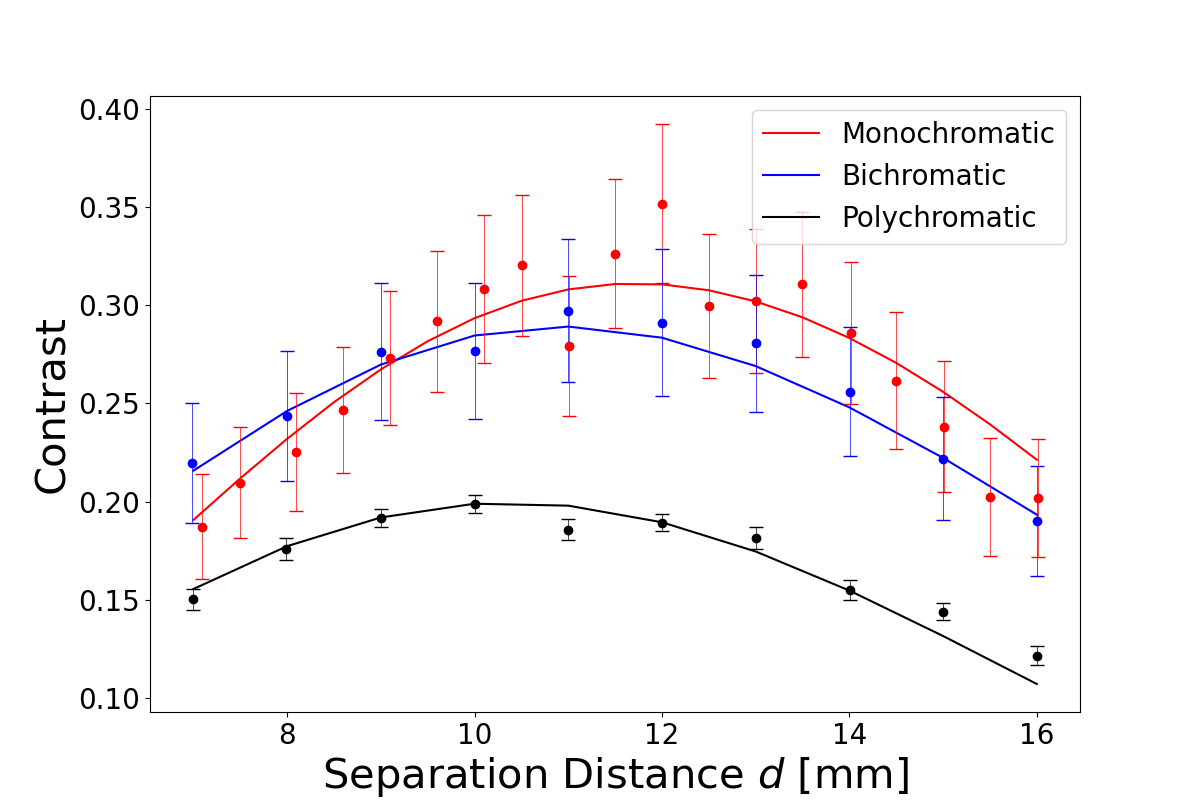}
    \caption{The experimentally measured contrast data as a function of grating separation $(d)$ reported for the three independent setups in Ref.~\cite{twogratings}, and the corresponding simulations (lines) done with the k-space model. Data uncertainties shown are purely statistical. The simulations were computed directly with the reported setup parameters and a least-squares fit for the slit size was performed. Good agreement is shown between the simulation and the measurements. All of the PGMI parameters for the three setups are shown in Table.~\ref{table:1}.
    }
    \label{fig:2PGMIexperiment}
\end{figure*}

The comparison between the experimentally measured contrast vs grating separation $(d)$ in these three setups and the simulated contrast using the k-space model is shown in Fig.~\ref{fig:2PGMIexperiment}. The simulations are done with a least-squares fit on the slit size to obtain the following best fit values: $230~\mu$m $\pm24~\mu$m for Monochromatic and Bichromatic setups, and $590~\mu$m $\pm46~\mu$m for the Polychromatic setup. Good agreement is found between the measured data and the simulation.

\begin{table}
\begin{tabular}{ |c|c|c|c|c|c|c|c|c|c|c| } 
\hline
\phantom{asd} & $L_1$ [m] & $L$ [m] & $p_{G}$ [$\mu$m] & $h$ [$\mu$m] & Res [$\mu$m] & $\lambda$ [\AA] & Slit [$\mu$m] & Slit fit [$\mu$m] \\ 
\hline
Mono & 1.2  & 3.04 & 2.4 & 9.3 & 100 & $4.4$ & 200 & 230$\pm$24\\ 
\hline
Bi & 1.73  & 3.51 & 2.4 & 9.3 & 100 & 24\% $2.2$ \& 76\% $4.4$& 200 & 230$\pm$24\\ 
\hline
Poly & 4.65  & 8.36 & 2.4 & 6.1 & 150 & Fig.~\ref{fig:wavelengthdist}a& 500 & 590$\pm$46\\ 
\hline
\end{tabular}
\caption{Parameters from the three independent two-PGMI setups that were characterized in Ref.~\cite{twogratings}. The ``slit fit'' is obtained using a least-squares fit on the slit size to obtain the simulated curves shown in Fig.~\ref{fig:2PGMIexperiment}. }
\label{table:1}
\end{table}

\subsection{Three-PGMI analysis}

\begin{figure*}
    \centering\includegraphics[width=\linewidth]{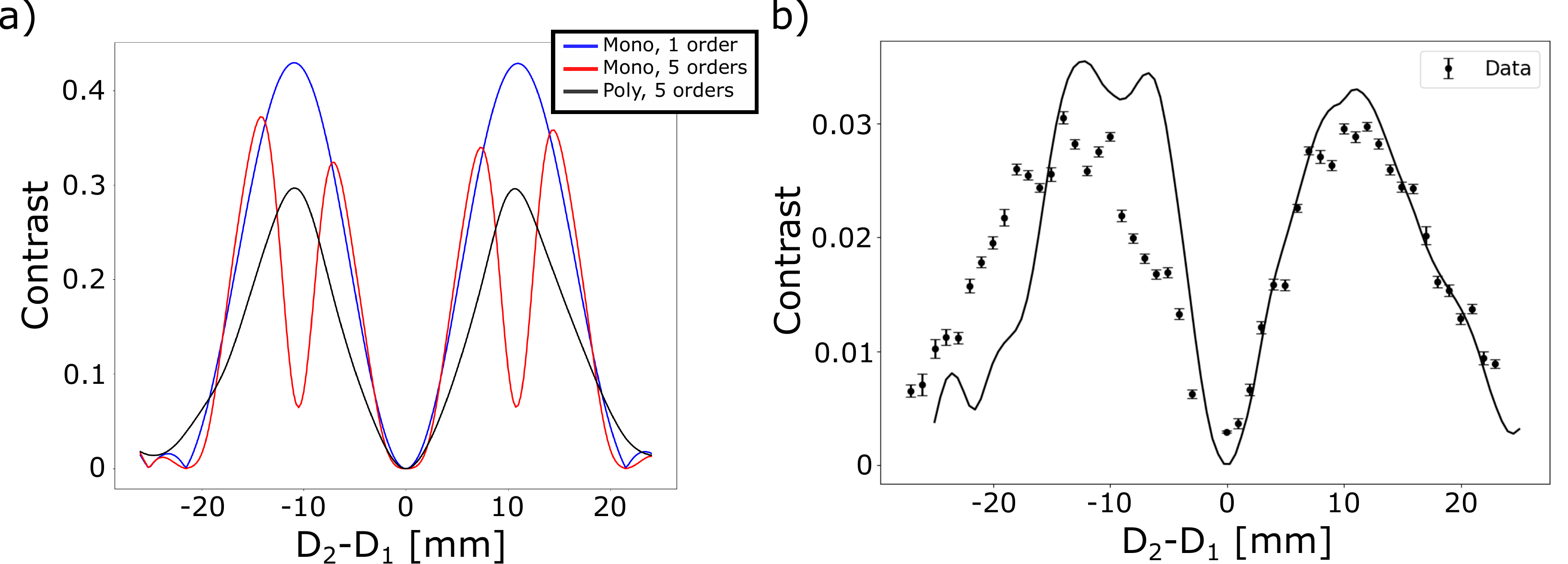}
    \caption{a) Simulated contrast for a three-PGMI setup with setup parameters reported as in Ref.~\cite{sarenac2018three}, but with considering ideal grating shape and alignment. Shown is the case of a monochromatic beam with $\lambda=5$~\AA~and with the first \& fifth (blue \& red) diffraction order approximation. The change in shape with the contrast dip is important to note if working with a monochromatic, three-PGMI setup. The black curve is the simulation that considers the polychromatic wavelength distribution of Ref.~\cite{sarenac2018three} (see Fig.~\ref{fig:wavelengthdist}a) and the fifth diffraction order approximation. Therefore, the measured contrast of $\approx3~\%$ of Ref.~\cite{sarenac2018three} was well below the expected maximum contrast of $\approx30~\%$. b) The middle grating possessed a high aspect ratio and baseball bat-like shape (see figure 2 of Ref.~\cite{heacock2019angular}) thereby increasing the influence of grating rotations about the vertical axis. If we consider the additional degree of freedom of grating rotations a good estimate for the observed contrast can be obtained. Shown is one such example where the first and third grating were rotated by $3.5\degree$ and the middle grating by $4.4\degree$. These misalignments are well within the possible experiment setup errors.
    }
    \label{fig:3PGMIexperiment}
\end{figure*}

Given the versatility of the model we can easily explore the three-PGMI setup that was reported in Ref.~\cite{sarenac2018three}. Let us consider the setup parameters of Ref.~\cite{sarenac2018three}: grating period $p_G=2.4~\mu$m, first and third grating height $h=16~\mu$m and middle grating height of $h=30~\mu$m, distance from the slit to the first grating $L_s=4.704~$m, distance from first to middle grating $L_1=4.6~$cm, slit size $500~\mu$m, and a camera pixel size $150~\mu$m. Exploring the contrast for the given parameters with the k-space model shows several interesting features as shown on Fig.~\ref{fig:3PGMIexperiment}a. First of all, if we consider a monochromatic beam at $5$~\AA~and the first diffraction order approximation then the contrast profile has the familiar shape and reaches a maximum of $40~\%$. However, if we include the coherence up to the fifth order, the contrast profile changes and shows a significant dip near the area of the expected maximum contrast. Further modeling can show that the location of the dip varies with the relative position of the middle grating to the middle of the setup $L/2$. This is important to note that if working with a monochromatic, three-PGMI setup as the first order diffraction approximation is not adequate to describe the behaviour. Lastly, if we include the polychromatic wavelength distribution of Fig.~\ref{fig:wavelengthdist}a, then the contrast profile is again of the familiar shape even with the higher orders included. 

The reported contrast in Ref.~\cite{sarenac2018three} was 3~\% and it was surprisingly lower than the expected peak contrast of $\approx32~\%$. Ref.~\cite{heacock2019angular} suggested that the cause of the problem might have been the vertical rotational alignment of the middle grating which possesses a high aspect ratio. The analysis was centered on analyzing the magnitude of the grating diffraction orders as a function of grating rotation, and extrapolating this change to be a change in maximum obtainable contrast. Here we can compute and explore the behaviour of the contrast profile directly. Projecting the SEM profile of the middle grating (see figure 2 of Ref.~\cite{heacock2019angular}) along different angles we can compute the effective profiles that the neutron would see when transversing the rotated grating~\cite{heacock2020neutron}. Plugging the resulting profile into the model (see postulate 2 in Section III) allows us to simulate the entire system. If we consider the additional degree of freedom of grating rotations a good estimate for the observed contrast can be obtained. Shown on Fig.~\ref{fig:3PGMIexperiment}b is one such example where the first and third grating were rotated by $3.5\degree$ and the middle grating by $4.4\degree$. These misalignments are well within the possible experiment setup errors. The simulation was done with a consideration of diffraction orders spanning from -5 to 5 in steps of 0.1. This was necessary because the diffraction spectrum of the rotated grating contains appreciable amplitudes at non-integer values of diffraction orders.

\subsection{Sample characterization with dark field imaging}

\begin{figure*}
    \centering\includegraphics[width=\linewidth]{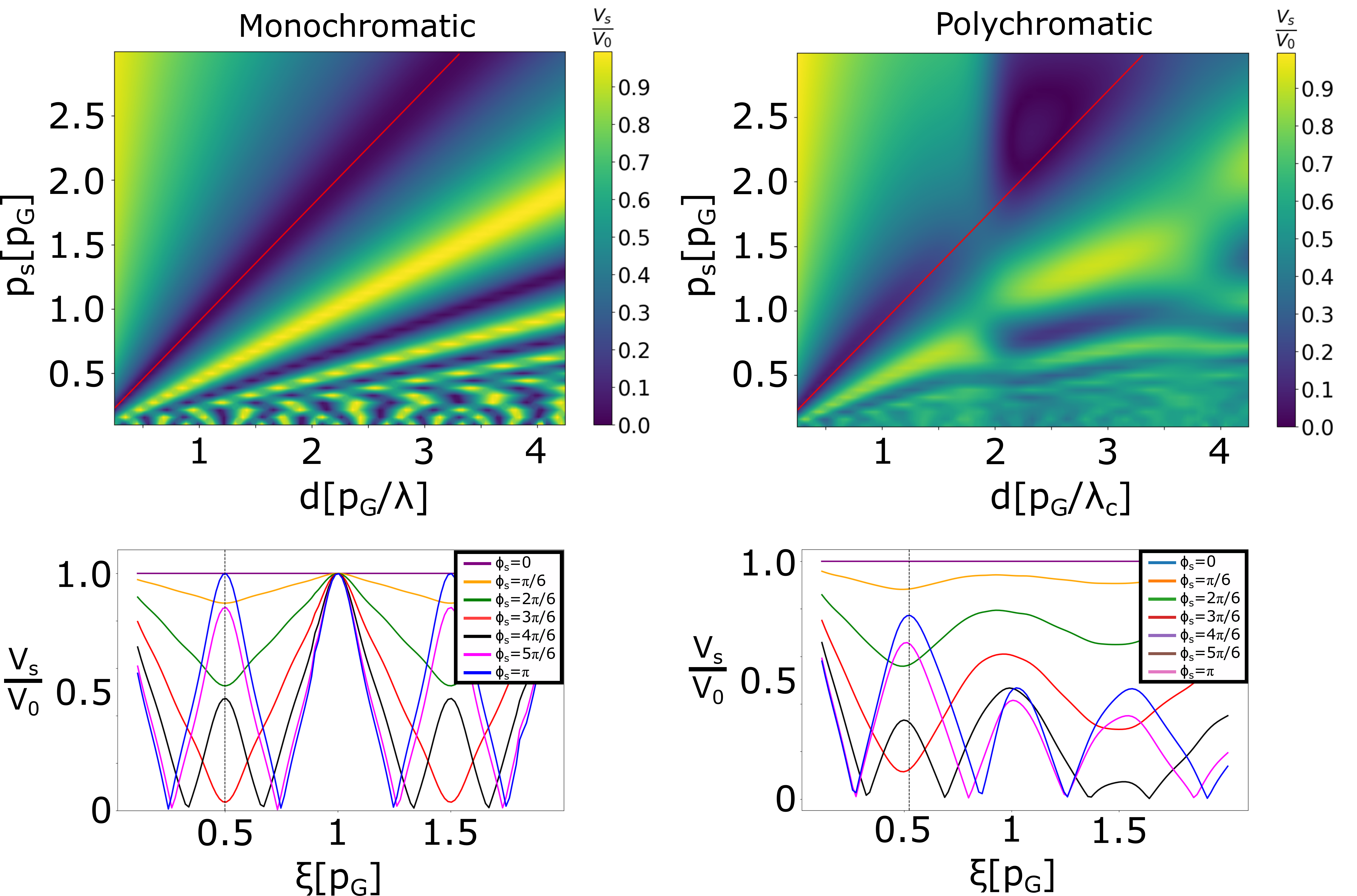}
    \caption{The simulated two-PGMI DFI signal when considering an ideal phase-grating of period $p_s$ and phase $\phi_s$ as the sample. DFI$=V_s/V_0$ where $V_s$ ($V_0$) is the contrast with (without) the sample. The left column is for the monochromatic case and the right column is for the polychromatic case of Fig.~\ref{fig:wavelengthdist}a with $\lambda_c=5$~\AA~. The top row shows the DFI dependence on $p_s$ and grating-separation $d$ when $\phi_s=\pi/2$ and $L_s\approx L/2$. The red line along $p_s=2L_s\lambda d/L p_G$ shows the location of the highest DFI sensitivity (largest contrast drop). The bottom row shows the dependence of the DFI signal on the interferometer autocorrelation length $\xi$ (see Eq.~\ref{eqn:AC}) for various sample phases $\phi_s$ when $p_s=p_G$ and $L_s\approx L/2$. The monochromatic case nicely depicts how the DFI signal (triangular wave) is the real space correlation function of the sample (rectangular grating).
    }
    \label{fig:DFIgrating}
\end{figure*}

Grating interferometers can be used to measure the microstructure of samples via a technique called Dark Field Imaging (DFI). Similar to spin-echo small-angle neutron scattering (SESANS), it has been shown that in a grating interferometer the ratio of contrast with the sample ($V_s$) to the contrast without the sample ($V_0$) is directly proportional to the real space correlation function of the sample $(C[\xi])$~\cite{andersson2008analysis,lynch2011interpretation,strobl2014general}:

\begin{align}
    \text{DFI}(\xi)=\frac{V_s(\xi)}{V_0(\xi)}\propto C(\xi)
    \label{eqn:generalDFI}
\end{align}

\noindent where:

\begin{align}
    \xi=\frac{\lambda L_s}{p_M}=\frac{\lambda L_s d}{p_GL}
    \label{eqn:AC}
\end{align}

\noindent is referred to as the autocorrelation length of the interferometer, $L_s$ is the distance of the sample to the detector if the sample is downstream of the gratings, or the distance to the source if the sample is upstream of the gratings. The DFI measurements therefore involve obtaining the two contrast values as $\xi$ is varied, either by varying the phase-grating separation distance $d$ or the distance between the sample and the detector $L_s$.

With the k-space model we can easily analyze the effect of a sample on the observed contrast for a wide range of parameters. Let us consider the sample to be an ideal 50~\% duty cycle phase-grating with period $p_s$ and phase $\phi_s=Nb_c\lambda h$:

\begin{align}
	f_s(x)=\frac{\phi_s}{2} \text{sgn}\left(\text{sin}\left[\frac{2\pi }{p_s}x\right]\right).
	\label{eqn:gratingsampleProfile}
\end{align}

\iffalse
\begin{figure*}
    \centering\includegraphics[width=\linewidth]{DFIgrating.pdf}
    \caption{The simulated two-PGMI DFI signal ("contrast with sample"/"contrast without sample") when considering a dilute solutions of monodisperse spheres with radius $R$ as the sample. The left column is for the monochromatic case and the right case is for the polychromatic case of Fig.~\ref{fig:wavelengthdist}a. The top row shows the DFI dependence on $R$ and grating-separation $d$. The bottom row shows the dependence of the DFI signal on the interferometer autocorrelation length $\xi$ (where $\xi=\lambda L_s d/p_GL$) for various $R$. The monochromatic case demonstrates how the DFI signal (triangular wave) is the real space correlation function of the sample (rectangular grating).
    }
    \label{fig:DFIsphere}
\end{figure*}
\fi

\noindent The top row of Fig.~\ref{fig:DFIgrating} shows the DFI signal $(V_s/V_0)$ as a function of $p_s$ and grating separation $d$, for sample phase of $\phi_s=\pi/2$ and for both the monochromatic and polychromatic cases. Here we can note that the largest contrast drop is obtained for $p_s=p_G$ when the sample is placed near the middle of the setup $L_s\approx L/2$ and the two-PGMI is in the optimal contrast condition of $d=p^2_G/\lambda$. Therefore, the two-PGMI typically probes sample periodicity on the order of the grating period (micrometers) similar to a one grating Talbot-Lau grating interferometer~\cite{strobl2014general}, but with the added benefit that the observed contrast is in the moir\'e regime (millimeters).

The bottom row of Fig.~\ref{fig:DFIgrating} shows the DFI signal for $p_s=p_G$ and various sample phases $\phi_s$. The x-axis is redefined from $d$ to $\xi$ according to Eq.~\ref{eqn:AC}. From the monochromatic case it is evident that the DFI signal (triangular wave) is the autocorrelation function of the sample profile (square wave) for any $p_s$. We can derive the equation for the expected DFI signal to be:

\begin{align}
	\text{DFI}(\xi)\approx\left|1+\frac{2\phi_s}{\pi}(C[\xi]-1)\right|
	\label{eqn:DFI}
\end{align}

\noindent As seen in the bottom row of Fig.~\ref{fig:DFIgrating}, a higher order harmonic appears for $\phi_s>\pi/2$ at the location of maximum contrast drop $\xi=p_s/2$. This behaviour has been observed with the inverse Talbot-Lau setup~\cite{kim2022analysis}. The polychromatic case is somewhat different because although each wavelength would produce the autocorrelation function of the sample profile, the amount that each wavelength contributes to the total observed intensity is different and can not be decoupled without a forward propagation simulation. In order to determine the sample autocorrelation function from polychromatic data one could simply perform the k-space simulation for the exact experimental parameters.

\begin{figure}
    \centering\includegraphics[width=\linewidth]{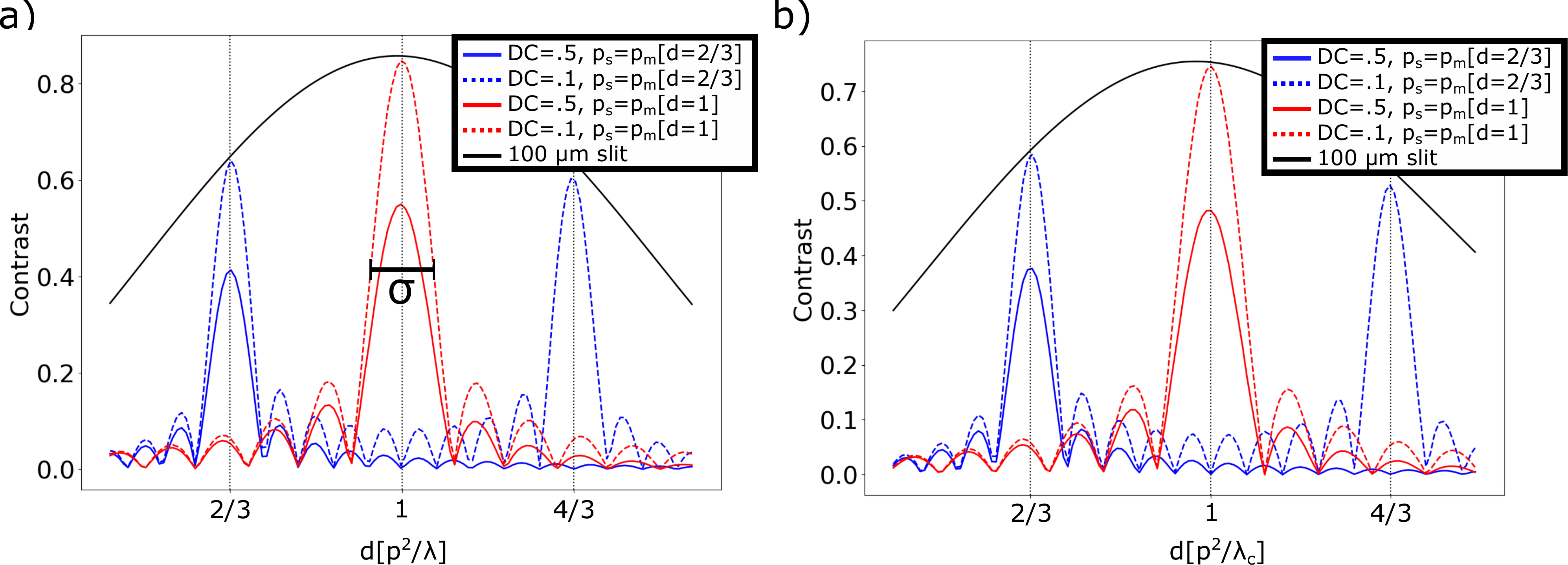}
    \caption{The source grating acts as a notch filter selecting out contrast for the case where the frequency of the source grating ($2\pi/p_s$) equals a multiple of the observed moir\'e frequency ($2\pi/p_M$). Plotted are three cases of source grating duty cycle (DC) that demonstrate the effect on filtered contrast and decreased ability to filter higher harmonics with increased duty cycle. a) shows the case for a monochromatic beam and b) for a polychromatic wavelength distribution of Fig.~\ref{fig:wavelengthdist}a with $\lambda_c=5$~\AA. The width $(\sigma)$ of the spectrum with the source grating is inversely proportional to the beam size at the source grating. The shown example is for the consideration of an input beam covering 10 source grating periods.
    }
    \label{fig:sourceGrating}
\end{figure}

\subsection{Source Grating}

One way to increase the number of neutrons that reach the camera is to use an absorption grating as a ``source grating'', which is essentially an array of slits. Simulating the source grating with the k-model can be done through a convolution with the source grating profile in place of the slit as done before, and the outcome is illustrated in Fig~\ref{fig:sourceGrating}. The source grating acts as a notch filter selecting out the contrast for the case where the frequency of the source grating ($2\pi/p_s$) equals a multiple of the observed moir\'e frequency ($2\pi/p_M$). It can be observed that the contrast is suppressed for $p_M\neq n p_s$, while the duty cycle of the source grating determines the amplitude of the max contrast and the presence of the higher order peaks. Lastly, we can note that the width $(\sigma)$ of the spectrum with the source grating is inversely proportional to the beam size at the source grating.

\subsection{Maximum contrast conditions}

The source grating is a powerful addition to a PGMI setup due to the increased neutron flux. It follows that it would be desirable to increase the maximum contrast at each particular grating separation distance. From Fig.~\ref{fig:TalbotandMoireIntensity}b it is apparent that maximal contrast of the two-PGMI occurs at $L_2=z_{TM}/2$, where $L_2=L-d-L_1$ is the distance from the second grating to the camera and $z_{TM}$ is given by Eq.~\ref{eqn:zTM}. Therefore we can calculate the conditions for maximal contrast given the parameters of a particular setup. With a fixed L, the maximal contrast occurs for 

\begin{align}
    L-d-L_1=\frac{p^2L(d+L_1)}{\lambda d(d+2L_1)}.
\end{align}

\noindent With the approximation of $L_1>>d$ we get:

\begin{align}
    d=\frac{Lp^2}{2\lambda(L-L_1)}\approx\frac{p^2}{\lambda}=\frac{z_T}{2}
    \label{eqn:optimald}
\end{align}

\noindent where for the approximation on the right we used $L_2= L_1$. In other words, in order for the camera in the two-PGMI setup to be located at the optimal contrast location of $\nicefrac{z_{TM}}{2}$ when the gratings are in the middle of the setup, the second grating needs to at the half-Talbot distance $(\nicefrac{z_T}{2})$ of the first grating.

\begin{figure}
    \centering\includegraphics[width=\linewidth]{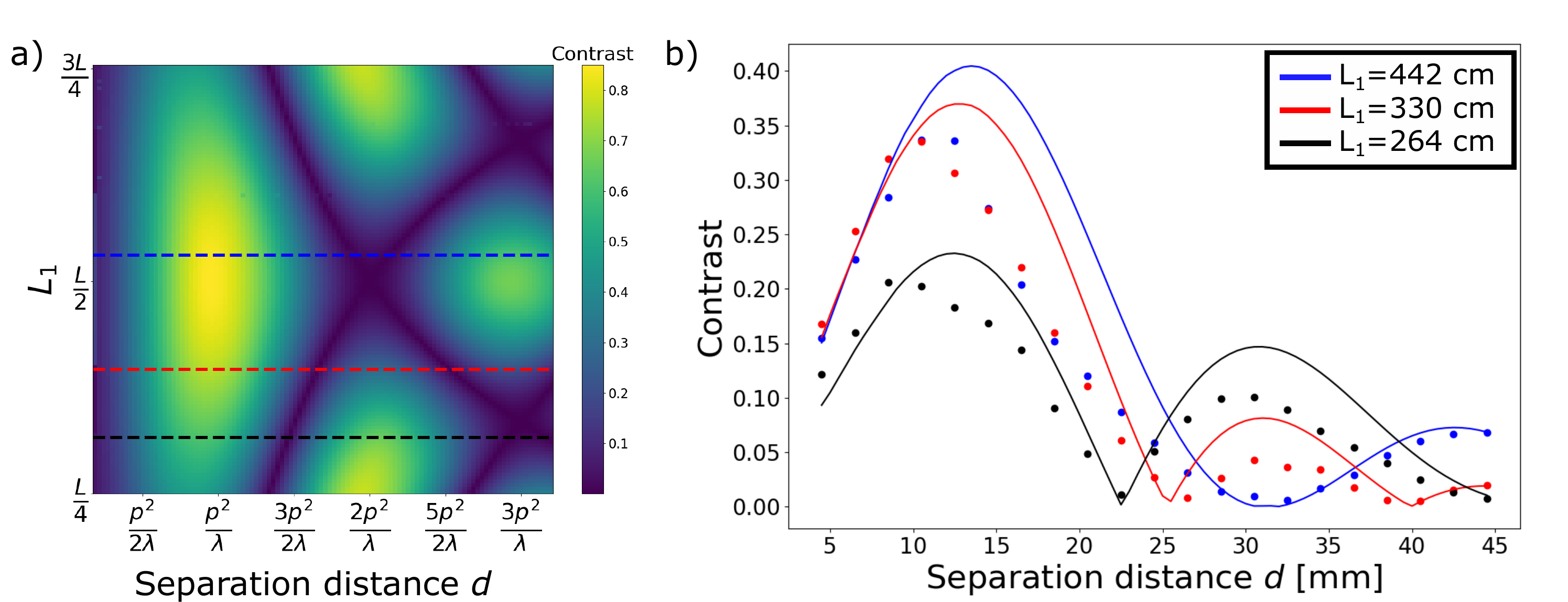}
    \caption{a) Ideal contrast behaviour as the distance from the slit to the first grating ($L_1$) is varied. b) To test this prediction an experiment was performed with a pulsed neutron beam and setup parameters: grating period $p_G=3~\mu$m, grating height $h=8.53~\mu$m, camera resolution of $100~\mu$m, and the wavelength distribution is shown on Fig.~\ref{fig:wavelengthdist}b, where the time of flight was set to select the $5-6$~\AA~wavelength range indicated. Note that the phase-gratings with $h=8.53~\mu$m were fabricated to act as $\pi/2$ phase-gratings for 9~\AA~neutrons. The model fit parameters that were optimized are $L=8.36$m, slit size $450~\mu$m, and $L_1=\{4.42~m,3.30~m,2.64~m\}$ for the three cases. 
    }
    \label{fig:contrastLines}
\end{figure}

From Eq.~\ref{eqn:zTM} we can note that the moir\'e period at the fixed camera is independent of the location of the gratings in the setup. Therefore, when the camera stays at a constant distance away from the source, as is the case in typical setups, as $d$ is varied the contrast will oscillate according to camera's location relative to $\nicefrac{z_{TM}}{2}$. One approach to maintain higher contrast is to vary $L_1$ with $d$:

\begin{align}
    L_1%=\frac{L (-p^2 + 2 d \lambda)}{2 d \lambda}=
    =L \left(1-\frac{p^2}{2 d \lambda}\right) =L\left(1 -\frac{z_T}{4 d}\right) 
    \label{eqn:optimalLc}
\end{align}

Using the k-space model we can explore this idea to determine several interesting conclusions. Fig~\ref{fig:contrastLines}a) shows the contrast behaviour as a function of $L_1$ and $d$. To test this prediction an experiment was performed with a pulsed neutron beam at the Energy-Resolved Neutron Imaging System (RADEN)~\cite{shinohara2020energy}, located at beam line BL22 of the Japan Proton Accelerator Research Complex (J-PARC) Materials and Life Science Experimental Facility (MLF). The setup parameters were: grating period $p_G=3~\mu$m, grating height $h=8.53~\mu$m, camera resolution of $100~\mu$m, and the wavelength distribution is shown on Fig.~\ref{fig:wavelengthdist}b, where the time of flight was set to select out the $5-6$~\AA~wavelength range indicated. Note that the phase-gratings with $h=8.53~\mu$m were optimized to act as $\pi/2$ phase-gratings for 9~\AA~neutrons. The model fit parameters that were optimized are $L=8.36~$m, slit size $450~\mu$m, and $L_1=\{4.42~m,3.30~m,2.64~m\}$ for the three cases. The experimental results along with a simulation are shown on Fig.~\ref{fig:contrastLines}b).

An interesting note that can be shown with the k-space model is that the dark bands of low contrast on Fig.~\ref{fig:contrastLines} are due to the accumulated phase between the two gratings. Hence, to achieve maximal contrast for every grating separation distance requires the removal of the phase evolution between the two gratings. This can be achieved with two gratings without a gap where the period of one grating is larger than that of the other grating~\cite{hidrovo2023neutron}. Whereas variable period gratings are easily achieved in optics with widely available spatial light modulators~\cite{curtis2002dynamic}, the analogue neutron devices are still in their infancy.

\section{Conclusion}

We have developed a toolbox for analyzing neutron interferometers illuminated by cone beams. The model postulates are developed through the generalization of the Fresnel Scaling theorem. This forward propagation model allows for analysis of PGMI intensity and contrast given a wide range of setup parameters, non-ideal considerations, phase structures, and attenuation structures.  

The model was used to simulate experiments with two-PGMI and three-PGMI setups. Good agreement was found, and the enabled optimization provided informative estimates for the parameters that were used in those experiments. Furthermore, several interesting conclusions were reached. For example, it was shown that the so called ``far-field'' interference of PGMIs is a manifestation of a moir\'e of the Talbot effects from the two phase-gratings. Furthermore, the model predicts an oscillation of the contrast as a function of the distance between the first phase-grating and the slit. This was experimentally tested and confirmed with a two-PGMI at the RADEN facility at the J-PARC.

The introduced model provides a backbone for future extensions, such as two-dimensional phase-gratings and characterization of samples with irregular structures. The model naturally enables the exploration of various sample structures and the expected dark-field imaging signal that they would produce. Furthermore, the addition of the orthogonal wave vectors and propagators can be accomplished in analogous fashion to the one-dimensional case presented here.

\section{Acknowledgements}

This work was supported by the Canadian Excellence Research Chairs (CERC) program, the Natural Sciences and Engineering Research Council of Canada (NSERC) Discovery program, Collaborative Research and Training Experience (CREATE) program, the Canada  First  Research  Excellence  Fund  (CFREF), and the National Institute of Standards and Technology (NIST) and the US Department of Energy, Office of Nuclear Physics, under Interagency Agreement 89243019SSC000025. The pulsed neutron experiment at J-PARC MLF was performed
under a user program (Proposal No. 2022A0104).

\bibliography{refs}

\clearpage

\section*{Appendix}

\subsection{Pseudo code example}

\begin{figure}[!htb]
    \centering\includegraphics[width=.75\linewidth]{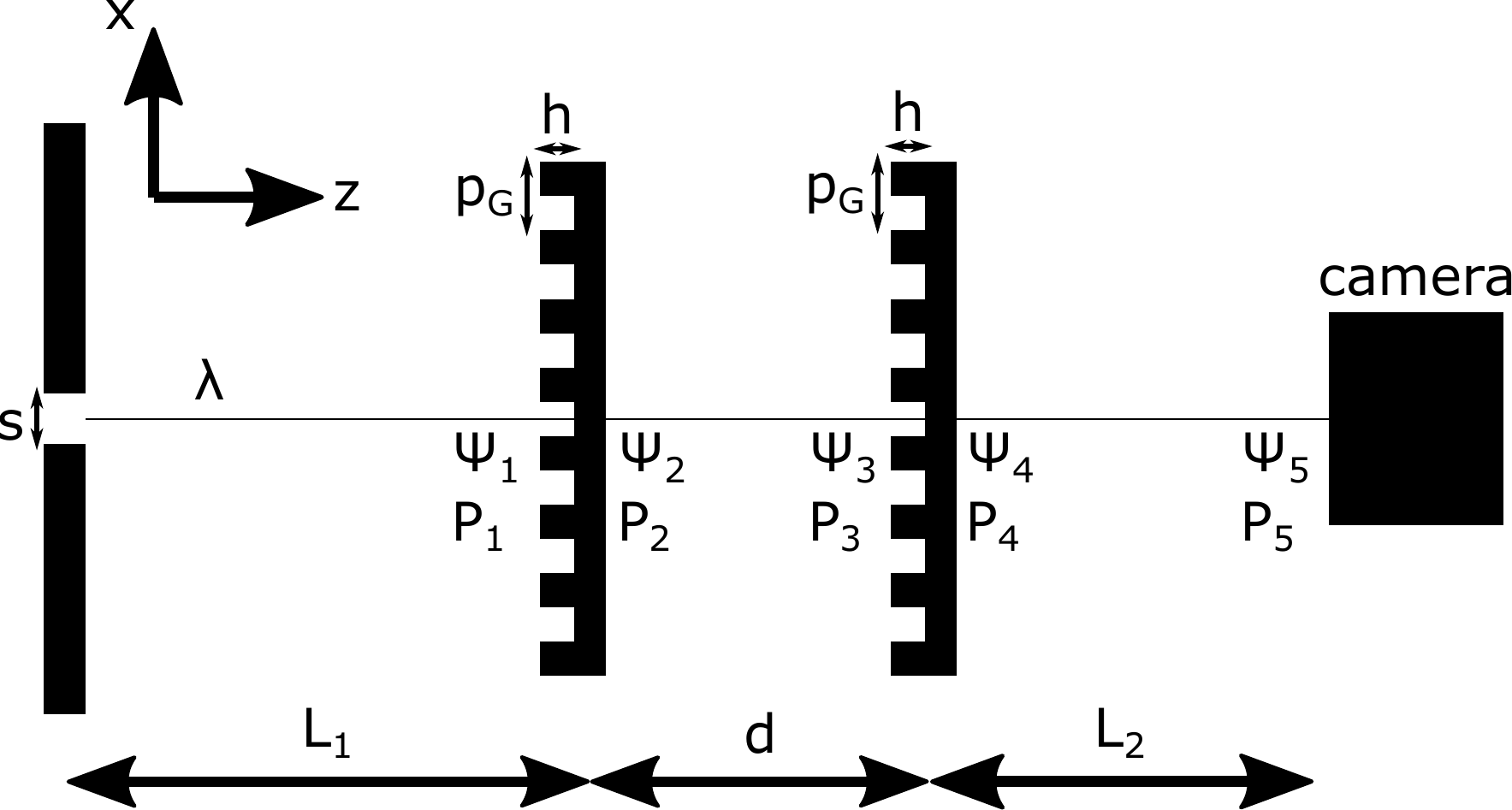}
    \caption{Two-PGMI setup for the pseudo code example.
    }
    \label{fig:pseudo}
\end{figure}

Here we provide an implementation example of the described k-space model by describing the pseudo code to compute contrast and intensity of a typical two-PGMI. Consider the setup of Fig.~\ref{fig:pseudo} with the following parameters: incoming neutrons with wavelength $\lambda$, equal period $p_G$ and height $h$ for the two phase-gratings, distance from the slit to the first phase-grating $L_1$, distance between the two phase-gratings $d$, distance from the second phase-grating to the camera $L_2$, slit size $s$, and camera pixel size $w$. 

The initial wavefunction is taken to be a plane wave with wave vector $k_0=2\pi/\lambda$ moving along the z-direction: $\Psi_1=\delta(k_x,k_z-k_0)$. For implementing the model it is convenient to use an array of two by one matrices where the first value of each matrix is the transverse wave vector and the second value is the complex amplitude associated with this particular wave vector. Therefore the initial array is: 

\begin{align}
	P_1 =\{\{0,1\}\}.
	\label{eqn:codePsi1}
\end{align}

Postulate 2 states that each new object with a spatial phase profile of $f(x)$ introduces a convolution between the incoming wavefunction and $\mathcal{F}\{e^{if(x)}\}$. After passing through the first phase-grating the convolution kernel is given by Eq.~\ref{eqn:gratingTerms}:

\begin{align}
	\mathcal{F}\{e^{-if(x)}\} = \delta(k_x)\cos\left(\frac{\alpha}{2}\right) 
	+  \sin\left(\frac{\alpha}{2}\right) \sum_{m} \frac{2e^{-ik_Gmx_1}}{\pi m}\delta\left(k_x-mk_G\right),
	\label{eqn:gratingTerms21}
\end{align}

\noindent where $x_1$ is the transverse translation of the phase-grating which is typically modulated to achieve phase stepping. Let us further simplify the example to consider $\pi/2$ phase-gratings, that is $\alpha=Nb_c\lambda h=\pi/2$. Then after the convolution the array becomes:

\begin{align}
	P_2 =\sum_m\{\{mk_G,a_m\}\}=\bigg\{...\bigg\{-k_G,-\frac{\sqrt{2}e^{ik_Gx_1}}{\pi}\bigg\},\bigg\{0,\frac{1}{\sqrt{2}}\bigg\},\bigg\{k_G,\frac{\sqrt{2}e^{-ik_Gx_1}}{\pi}\bigg\}...\bigg\},
	\label{eqn:codePsi2}
\end{align}

Postulate 1 tells us that the propagation from the first grating to the second grating scales the transverse wave vectors as $k_x\rightarrow k_x/M_1$ where $M_1=1+d/L_1$. Postulate 3 tells us that the amplitude of each magnified wave vector picks up a phase of $\text{Exp}[idk_z/M_1]\approx\text{Exp}[idk_0(1-k_x^2/[2k_0^2])/M_1]$. We can factor out the common phase of $\text{Exp}[idk_0/M_1]$ to get the reduced phase with propagation $\text{Exp}[-idk_x^2/(2k_0M_1)]$. Therefore our updated array becomes:

\begin{align}
	P_3 =\bigg\{\bigg\{-\frac{k_G}{M_1},-\frac{\sqrt{2}}{\pi}e^{-idK_z\left[-\frac{k_G}{M_1},\frac{k_0}{M_1}\right]}\bigg\},\bigg\{0,\frac{1}{\sqrt{2}}\bigg\},\bigg\{\frac{k_G}{M_1},\frac{\sqrt{2}}{\pi}e^{-idK_z\left[\frac{k_G}{M_1},\frac{k_0}{M_1}\right]}\bigg\}\bigg\},
	\label{eqn:codePsi3}
\end{align}

\noindent where for clarity we have set $x_1=0$, displayed only the $m=-1,0,1$ diffraction orders, and defined the function for the reduced longitudinal wave vector:

\begin{align}
	K_z[k_x,k_0] =k_x^2/(2k_0)
	\label{reducedkz}
\end{align}

\noindent Similar to the first phase-grating, the second phase-grating does the convolution operation between the incoming wavefunction and Eq.~\ref{eqn:gratingTerms21}: 

\begin{align}
\nonumber
P_4 =&\bigg\{\bigg\{-\frac{k_G}{M_1}-k_G,\frac{2}{\pi^2}e^{-idK_z\left[-\frac{k_G}{M_1},\frac{k_0}{M_1}\right]}\bigg\},\bigg\{-k_G,-\frac{1}{\pi}\bigg\},\bigg\{\frac{k_G}{M_1}-k_G,-\frac{2}{\pi^2}e^{-idK_z\left[\frac{k_G}{M_1},\frac{k_0}{M_1}\right]}\bigg\},\\\nonumber
 &\bigg\{-\frac{k_G}{M_1},-\frac{1}{\pi}e^{-idK_z\left[-\frac{k_G}{M_1},\frac{k_0}{M_1}\right]}\bigg\},\bigg\{0,\frac{1}{2}\bigg\},\bigg\{\frac{k_G}{M_1},\frac{1}{\pi}e^{-idK_z\left[\frac{k_G}{M_1},\frac{k_0}{M_1}\right]}\bigg\},\\
  &\bigg\{-\frac{k_G}{M_1}+k_G,-\frac{2}{\pi^2}e^{-idK_z\left[-\frac{k_G}{M_1},\frac{k_0}{M_1}\right]}\bigg\},\bigg\{k_G,\frac{1}{\pi}\bigg\},\bigg\{\frac{k_G}{M_1}+k_G,\frac{2}{\pi^2}e^{-idK_z\left[\frac{k_G}{M_1},\frac{k_0}{M_1}\right]}\bigg\}\bigg\}.
    \end{align}

Similar to before, the propagation from the second grating to the camera scales the transverse wave vectors as $k_x\rightarrow k_x/M_2$ where $M_2=1+L_2/(L_1+d)$. Likewise the reduced phase with propagation is $\text{Exp}\left[-iL_2K_z\left[\frac{k_x}{M_2},\frac{k_0}{M_1M_2}\right]\right]$, where $k_x=mk_G/M_1$. However, given that $z=0$ was redefined to start at the second grating we need to add an additional scaling term of $M_1$ (see postulate 4) and get $\text{Exp}\left[-iL_2K_z\left[\frac{k_x}{M_2},\frac{k_0}{M_1M_2}\right]/M_1\right]$. The array then becomes:

\begin{align}
    P_5 =\bigg\{&\bigg\{-\frac{k_G(1+M_1)}{M_1M_2},\frac{2}{\pi^2}e^{-idK_z\left[-\frac{k_G}{M_1},\frac{k_0}{M_1}\right]}e^{-iL_2K_z\left[-\frac{k_G(1+M_1)}{M_1M_2},\frac{k_0}{M_1M_2}\right]/M_1}\bigg\},\\\nonumber
    &\bigg\{\frac{-k_G}{M_2},-\frac{1}{\pi}e^{-iL_2K_z\left[\frac{-k_G}{M_2},\frac{k_0}{M_1M_2}\right]/M_1}\bigg\},\\\nonumber
    &\bigg\{\frac{k_G(1-M_1)}{M_1M_2},-\frac{2}{\pi^2}e^{-idK_z\left[\frac{k_G}{M_1},\frac{k_0}{M_1}\right]}e^{-iL_2K_z\left[\frac{k_G(1-M_1)}{M_1M_2},\frac{k_0}{M_1M_2}\right]/M_1}\bigg\}\\\nonumber
     &\bigg\{-\frac{k_G}{M_1M_2},-\frac{1}{\pi}e^{-idK_z\left[\frac{k_G}{M_1},\frac{k_0}{M_1}\right]}e^{-iL_2K_z\left[-\frac{k_G}{M_1M_2},\frac{k_0}{M_1M_2}\right]/M_1}\bigg\},\\\nonumber
     &\bigg\{0,\frac{1}{2}\bigg\},\\\nonumber
     &\bigg\{\frac{k_G}{M_1M_2},\frac{1}{\pi}e^{-idK_z\left[\frac{k_G}{M_1},\frac{k_0}{M_1}\right]}e^{-iL_2K_z\left[\frac{k_G}{M_1M_2},\frac{k_0}{M_1M_2}\right]/M_1}\bigg\},\\\nonumber
     &\bigg\{\frac{k_G(M_1-1)}{M_1M_2},-\frac{2}{\pi^2}e^{-idK_z\left[\frac{k_G}{M_1},\frac{k_0}{M_1}\right]}e^{-iL_2K_z\left[\frac{k_G(M_1-1)}{M_1M_2},\frac{k_0}{M_1M_2}\right]/M_1}\bigg\},\\\nonumber
     &\bigg\{\frac{k_G}{M_2},\frac{1}{\pi}e^{-iL_2K_z\left[\frac{k_G}{M_2},\frac{k_0}{M_1M_2}\right]/M_1}\bigg\},\\\nonumber
     &\bigg\{\frac{k_G(M_1+1)}{M_1M_2},\frac{2}{\pi^2}e^{-idK_z\left[\frac{k_G}{M_1},\frac{k_0}{M_1}\right]}e^{-iL_2K_z\left[\frac{k_G(1+M_1)}{M_1M_2},\frac{k_0}{M_1M_2}\right]/M_1}\bigg\}\bigg\},\nonumber
	\label{eqn:codePsi5}
\end{align}

\noindent This array contains all of the information about the wavefunction at the camera: $\Psi_5(k_x)=\sum_n a_n \delta(k_x-k_n)$ where $k_n~(a_n)$  is the first (second) entry of each matrix in $P_5$. As per postulate 5, the intensity is given by $I(x)=\left|\mathcal{F}^{-1}\{\Psi(k_x)\}\right|^2$ where $I(x)$ was written out in Eq.~\ref{eqn:2PGMIconeIntensity}.

As per postulate 6, to calculate the contrast of the moir\'e wave vector $k_m$, we need to compute $H(k_x)=\Psi_5(k_x)*\Psi^*_5(k_x)$. Following through with our array in the pseudo code we compute $H(k_x)=P_5(k_x)*P^*_5(k_x)$. Note that there will be multiple entries with the same wave vector as different combinations of wave vectors contribute to the interference at a specific frequency. Therefore we need reduce $H(k_x)$ to combine all equal wave vector entries. For example $H(k_x)=\{...\{k_i,a\},\{k_i,b\},\{k_j,c\},\{k_j,d\}...\}\rightarrow H(k_x)=\{...\{k_i,a+b\},\{k_j,c+d\}...\}$. The contrast for a particular wave vector $k_m$ is given by:

\begin{align}
    V(k_m)=\left|\frac{2H(km)}{H(0)}\right|
\end{align}

Let us say that we wish to integrate the effect of a rectangular slit of width ``s'' and pixel side ``w''. Then according to Eq.~\ref{eqn:contrastfit}

\begin{align}
    V(k_m)=\left|\frac{2H(km)\text{sinc}(sk_m/2)\text{sinc}(wk_m/2)}{H(0)}\right|
    \label{eqn:pseudoContrast2}
\end{align}

Lastly, let's say we want to incorporate a particular wavelength distribution. Then one simply needs to calculate $H(k_x)=\int d\lambda p(\lambda) P_5(k_x,\lambda)*P^*_5(k_x,\lambda)$ and follow through to  Eq.~\ref{eqn:pseudoContrast2}.

\subsection{Wavelength distributions}

\begin{figure}[!htb]
    \centering\includegraphics[width=\linewidth]{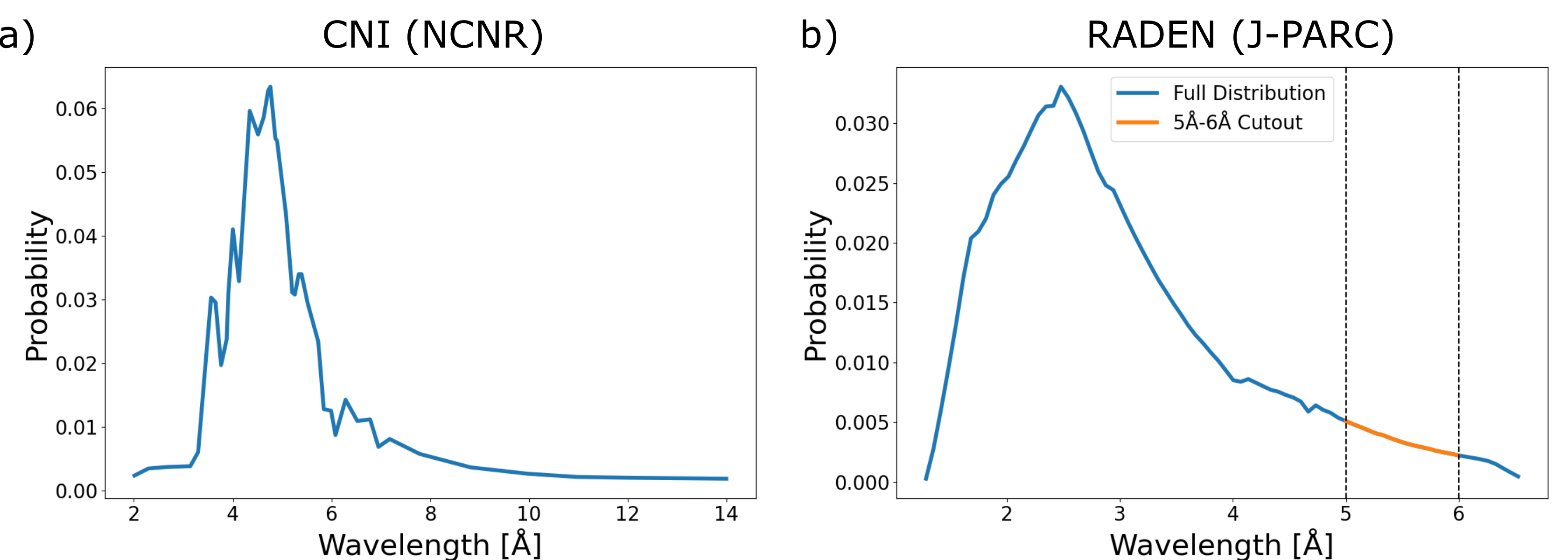}
    \caption{a) The wavelength distribution for the Cold Neutron Imaging (CNI) facility at the National Institute of Standards and Technology’s Center for Neutron Research (NCNR). This wavelength distribution was used for the simulation shown in Fig.~\ref{fig:3PGMIexperiment}b and the ``polychromatic case'' simulations of Figs.~\ref{fig:2PGMIexperiment},~\ref{fig:3PGMIexperiment}a,~\ref{fig:DFIgrating},~\ref{fig:sourceGrating}.
    b) The wavelength distribution for the RADEN facility at the Japan Proton Accelerator Research Complex (J-PARC). The indicated wavelength distribution cutout between 5~\AA~and 6~\AA~was used for the simulations shown in Fig.~\ref{fig:contrastLines}.
    }
    \label{fig:wavelengthdist}
\end{figure}

\end{document}